\documentclass[twocolumn,showpacs,preprintnumbers,amsmath,amssymb]{revtex4}

\usepackage{graphicx}
\usepackage{dcolumn}
\usepackage{bm}

\begin{document}


\title{Possibility of synthesizing doubly closed superheavy nucleus}

\author{Y.~Aritomo}
\affiliation{Flerov Laboratory of Nuclear Reactions, JINR, Dubna,
Russia }


\date{\today}

\begin{abstract}
The possibility of synthesizing a doubly magic superheavy nucleus,
$^{298}114_{184}$, is investigated on the basis of
fluctuation-dissipation dynamics. In order to synthesize this
nucleus, we must generate more neutron-rich compound nuclei because
of the neutron emissions from excited compound nuclei. The compound
nucleus $^{304}114$ has two advantages to achieving a high survival
probability. First, because of small neutron separation energy and
rapid cooling, the shell correction energy recovers quickly.
Secondly, owing to neutron emissions, the neutron number of the
nucleus approaches that of the double closed shell and the nucleus
obtains a large fission barrier. Because of these two effects, the
survival probability of $^{304}114$ does not decrease until the
excitation energy $E^{*}= 50$ MeV. These properties lead to a rather
high evaporation reside cross section.
\end{abstract}


\maketitle


The search for new elements is a long-standing important subject in
nuclear physics  \cite{ogan85,munz88}. According to
macroscopic-microscopic calculations \cite{myer66}, there should be
a magic island of stability surrounding the doubly magic superheavy
nucleus containing 114 protons and 184 neutrons. Attempts to
synthesize heavy elements with atomic numbers beyond $ Z \sim 100 $
have been active since the 1970s, making use of various developments
in experimental techniques \cite{ogan85,munz88,armb85}. For
superheavy elements around $Z \sim 114$ and $N \sim 184$, practical
combinations of a target and projectile, such as
$^{48}$Ca+$^{244}$Pu, have been used by the FLNR group
\cite{ogan99}. In this case, the neutron number of the compound
nucleus is less than $N=184$.

Actually, if we plan to synthesize the doubly magic superheavy
nucleus $^{298}114_{184}$, we must fabricate more neutron-rich
compound nuclei because of the neutron emissions from excited
compound nuclei. Since combinations of stable nuclei do not provide
such neutron-rich nuclei, the reaction mechanism for nuclei with $Z
=114, N > 184$ has rarely been investigated until now. However,
because of the characteristic properties of these nuclei, we find an
unexpected reaction mechanism for enhancing the evaporation residue
cross section. We report this mechanism in this paper.

As is well-known, in heavy systems around $Z \sim 80$, the
trajectory calculations with friction \cite{swia81,bloc86} were very
useful for the explanation of the extra- or extra-extra-push energy.
In superheavy mass region, however, the mean trajectory calculations
are not suitable, because mean trajectories cannot reach the
spherical shape region and around due to the strong dissipation
\cite{bjor82}. However, the extremely small part of distribution can
be found there due to fluctuation. Therefore, it is important to
take into account the fluctuating part from the mean trajectory. It
becomes necessary to solve a full dissipative dynamics, or a
fluctuation-dissipation dynamics with the Kramers (Fokker-Planck)
equation or with the Langevin equation \cite{agui89,wada93,toku99}.

Using the same procedure as described in reference \cite{ari04}, we
apply the fluctuation-dissipation model and employ the Langevin
equation for the fusion process. On the basis of our previous
studies \cite{arit97,arit99}, to investigate the fission process, we
employ the Smoluchowski equation which is a strong friction limit of
Fokker-Planck equation. Here, we take into account the
temperature-dependent shell correction energy.


The evaporation residue cross section $\sigma_{ER}$ is estimated as
\begin{equation}
\sigma_{ER} = \frac{\pi\hbar^2}{2{\mu_0}E_{cm}}\sum_{l=0}^{\infty}
(2l + 1)T_{l}(E_{cm},l)P_{CN}(E^{*},l)W(E^{*},l), \label{Xev}
\end{equation}
where $\mu_0$ denotes the reduced mass in the entrance channel.
$E_{cm}$ and $E^{*}$ denote the incident energy in the
center-of-mass frame and the excitation energy of the compound
nucleus, respectively. $E^{*}$ is given as $E^{*}=E_{cm}-Q$ with $Q$
denoting the $Q-$value of the reaction. $T_{l}(E_{cm},l)$ is the
capture probability of the $l$th partial wave, which is calculated
with the empirical coupled channel model \cite{zagr01}.
$P_{CN}(E^{*},l)$ is the probability of forming a compound nucleus
in competition with quasi-fission events. $W(E^{*},l)$ denotes the
survival probability of compound nuclei during de-exciting process.

To calculate $P_{CN}$, we employ the Langevin equation. We adopt the
three-dimensional nuclear deformation space given by two-center
parameterization \cite{maru72,sato78}.
The three collective parameters involved in the Langevin equation
are as follows: $z_{0}$ (distance between two potential centers),
$\delta$ (deformation of fragments) and $\alpha$ (mass asymmetry of
the colliding nuclei); $\alpha=(A_{1}-A_{2})/(A_{1}+A_{2})$, where
$A_{1}$ and $A_{2}$ denote the mass numbers of the target and
projectile, respectively. The detail is explained in reference
\cite{ari04}.

After the probabilities reaching the spherical shape and around, we
must treat extremely small probabilities in the decay process of the
compound nucleus. Therefore, we investigate the evolution of the
probability distribution $P(q,l;t)$ in the collective coordinate
space with the Smoluchowski equation \cite{arit97,arit99}. We employ
the one-dimensional Smoluchowski equation in the elongation degree
of freedom $z_{0}$, which is expressed as follows;

\begin{equation}
\frac{\partial}{\partial t}P(q,l;t)=\frac{1}{\mu\beta}
\frac{\partial}{\partial q} \left\{ \frac{\partial V(q,l;t)}
{\partial{q}}P(q,l;t) \right\} +
\frac{T}{\mu\beta}\frac{\partial^2}{\partial{q^2}}P(q,l;t)  .
\label{Smol}
\end{equation}
$q$ denotes the coordinate specified by $z_{0}$. $V(q,l;t)$ is the
potential energy, and the angular momentum of the system is
expressed by $l$. $\mu$ and $\beta$ are the inertia mass and the
reduced friction, respectively. For these quantities we use the same
values as in references \cite{arit97,arit99}.
 $T$ is the temperature of the compound nucleus calculated from
the excitation energy as $E^{*} = a T^2$ with $a$ denoting the level
density parameter of T\"oke and Swiatecki \cite{toke81}. The
temperature dependent shell correction energy is added to the
macroscopic potential energy,
\begin{equation}
V(q,l;t) = V_{\rm DM}(q)+\frac{\hbar^{2}l(l+1)}{2I(q)} +V_{\rm
shell}(q){\Phi}(t) , \label{Pot}
\end{equation}
where $I(q)$ is the moment of inertia of rigid body at coordinate
$q$. $V_{\rm DM}$ is the potential energy of the finite range
droplet model and $V_{\rm shell}$ is the shell correction energy at
$T=0$ \cite{ari04}.

The temperature dependence of the shell correction energy is
extracted from the free energy calculated with single particle
energies \cite{arit99,ohta95}. The temperature-dependent factor
${\Phi}(t)$ in Eq.~(\ref{Pot}) is parameterized as;
\begin{equation}
{\Phi}(t) = \exp{\left(-\frac{aT^2(t)}{E_d}\right)} ,
\end{equation}
following the work by Ignatyuk {\it et al.}\cite{igna75}. The
shell-damping energy $E_d$ is chosen as 20~MeV. The cooling curve
$T(t)$ is calculated by the statistical model code SIMDEC
\cite{arit99,ohta95}. We assume that the particle emissions in the
composite system are limited to neutron evaporation in the
neutron-rich heavy nuclei. When the temperature decreases as a
result of neutron evaporation, the potential energy $V(q,l;t)$
changes due to the restoration of shell correction energy.

The survival probability $W(E^{*}_{0},l;t)$ is defined as the
probability which is left inside the fission barrier in the decay
process;
\begin{equation}
W(E^{*}_{0},l;t)=\int_{\rm inside\ saddle} P(q,l;t)dq .
\end{equation}
Here, $E^{*}_{0}$ is the initial excitation energy of the compound
nucleus.

For the purpose of understanding well the characteristic enhancement
in the excitation function, we first discuss the evaporation residue
probability of one partial wave, i.e., of $l=10$, which is one of
the dominantly contributing partial waves \cite{arit99}.



In our previous studies \cite{arit97,arit99}, we showed the isotope
dependence of the evaporation reside cross section for $Z=114$. At
that time, we investigated the isotope dependence with $N \le 184$.
However, in order to synthesize the doubly magic nucleus
$^{298}114_{184}$ by hot fusion reactions, we must fabricate a more
neutron-rich compound nucleus of $N > 184$ because of the neutron
emissions from the excited compound nucleus. Taking into account the
neutron emissions, we investigate the possibility of synthesizing
$^{298}114_{184}$.

\begin{figure}
\centerline{
\includegraphics[height=0.50\textheight]{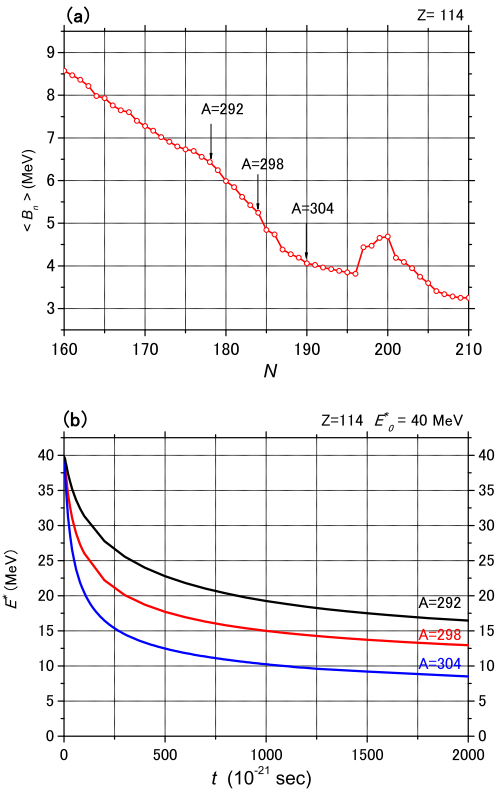}}
  \caption{(a)~Neutron separation energies averaged over four
successive neutron emissions $ \langle B_{n} \rangle$ for the
isotopes with $Z=114$ \cite{moll95}. (b)~Cooling curves of
$A=292,298$ and 304 with $Z=114$ at the initial excitation energy
$E^{*}_{0}=40$ MeV, that are derived by the statistical code SIMDEC
\cite{arit99,ohta95}.}
\end{figure}

The neutron separation energy depends on the neutron number.
Figure~1(a) shows the neutron separation energies averaged over four
successive neutron emissions $ \langle B_{n} \rangle$ for the
isotopes with $Z=114$. We use the mass table in reference
\cite{moll95}. $\langle B_{n} \rangle$s of $A=292,298$ and 304 are
6.43, 5.25 and 4.06 MeV, respectively. With increasing neutron
number of the nucleus, the neutron separation energy becomes small.
Therefore many neutrons evaporate easily from the neutron-rich
compound nuclei.

Because of rapid neutron emissions, the cooling speed of the
compound nucleus is very high. Figure~1(b) shows the cooling curves
of $A=292,298$ and 304 at the initial excitation energy
$E^{*}_{0}=40$ MeV, that were derived using the statistical code
SIMDEC \cite{arit99,ohta95}. In the case of $A=304$, the excited
compound nucleus cools rapidly and the fission barrier recovers at a
low excitation energy.

\begin{figure}
\centerline{
\includegraphics[height=0.50\textheight]{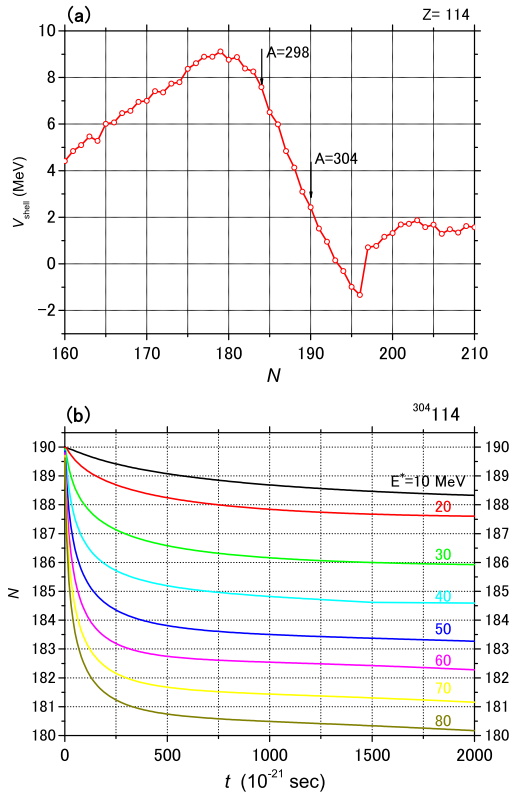}}
  \caption{(a)~Shell correction energies $V_{\rm shell}$ of
isotopes with $Z=114$ \cite{moll95}. (b)~Time evolution of the
neutron number for the de-exciting nucleus $^{304}114_{190}$ for
eight different initial excitation energies.}
\end{figure}

Moreover, owing to the neutron emissions, the neutron number of the
de-exciting nucleus with $A=304$ approaches that of a nucleus with
the double closed shell $Z=114, N=184$. Figure~2(a) shows the shell
correction energies $V_{\rm shell}$ of isotopes with $Z=114$
\cite{moll95}. $V_{\rm shell}$ of the $A=304$ $(N=190)$ nucleus is
smaller than that of the $A=298$ $(N=184)$ nucleus. However, in the
de-exciting process of the nucleus with $A=304$ $(N=190)$, the
neutron number approaches $N=184$ because of neutron emission. In
Fig.~2(b), the time evolution of the neutron number for the compound
nucleus $^{304}114_{190}$ is shown for eight different initial
excitation energies, as calculated by SIMDEC \cite{arit99,ohta95}.
At a high initial excitation energy, the neutron number of the
compound nucleus quickly approaches $N \sim 184$, which is that of a
neutron closed shell. This means the rapid appearance of a large
fission barrier.

The compound nucleus with $^{304}114$ has two advantages to
obtaining a high survival probability. First, because of small
neutron separation energy and rapid cooling, the shell correction
energy recovers quickly. Secondly, because of neutron emissions, the
number of neutrons in the nucleus approaches that in the double
closed shell, and a large shell correction energy is attained.

\begin{figure}
\centerline{
\includegraphics[height=0.50\textheight]{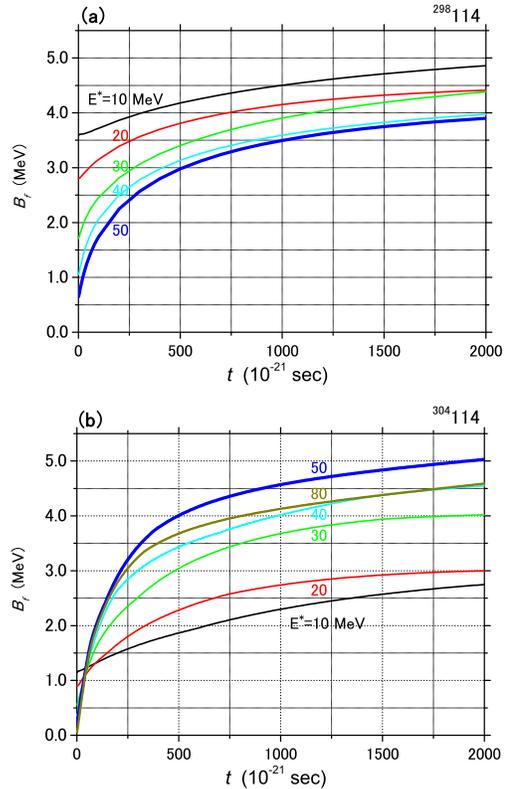}}
  \caption{Time evolution of the fission barrier height $B_{f}$ for the de-exciting
nuclei (a) $^{298}114$ and (b)$^{304}114$.}
\end{figure}

Generally, at a high excitation energy, the recovery of the shell
correction energy is delayed. On the other hand, at a low excitation
energy, the shell correction energy is established. Figure~3(a)
shows the time evolution of the fission barrier height $B_{f}$ for
$^{298}114$. We can see that the restoration of shell correction
energy is increasingly delayed with increasing excitation energy.
Using the Smoluchowski equation, we calculate the survival
probability, which is denoted by the red line in Fig.~4. With
increasing excitation energy, the survival probability decreases
drastically.

However, for $^{304}114$, the situation is opposite. At an
excitation energy of 50 MeV, the fission barrier recovers faster
than in the cases with lower excitation energies, as shown in
Fig.~3(b). The reason is the double effects, that is to say, the
rapid cooling and rapid approach to $N \sim 184$. The survival
probability of $^{304}114$ is denoted by the blue line in Fig.~4. It
is very interesting that the excitation function of the survival
probability has a flat region around $E^{*}=20 \sim 50$ MeV. At
$E^{*}=50$ MeV, the survival probability of $^{304}114$ is three
orders magnitude larger than that of $^{298}114$. For reference, the
survival probability of $^{300}114$ is denoted by the green line in
Fig.~4.

\begin{figure}
\centerline{
\includegraphics[height=0.40\textheight]{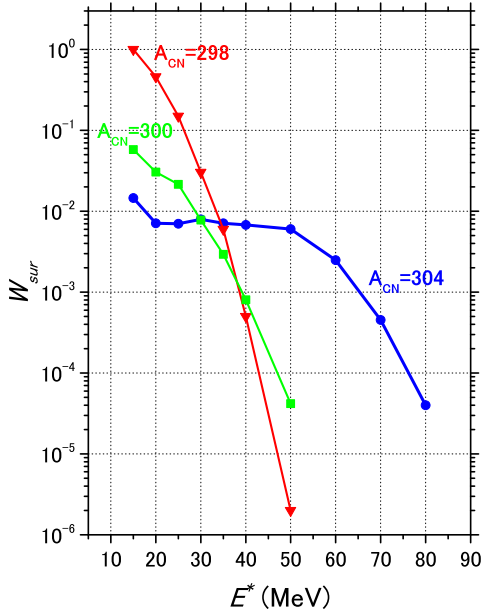}}
  \caption{Survival probabilities for $^{298}114, ^{300}114$ and
  $^{304}114$, which are calculated by the one-dimensional Smoluchowski equation.}
\end{figure}

For the ideal combinations for synthesizing the compound nucleus
$^{304}114$, the fusion probabilities for each system are shown in
Fig.~5, which were calculated using the Langevin equation, except
for the reaction $^{152}$La+$^{152}$La. This symmetric reaction
system with extremely small fusion probability is applied to the
one-dimensional Smoluchowski equation. The combinations of the
projectile and target are indicated in Fig.~5. The corresponding
Bass potential barriers are denoted by the arrows \cite{bass74}. We
show the fusion probabilities above the barrier, because we use the
classical models. To multiply the fusion probability by the survival
probability of $^{304}114$ in Fig.~4, we obtain the evaporation
residue cross section of superheavy elements, as shown in Fig.~6.
The cross section is rather high. It is expected that neutron-rich
isotopes are more favorable for the enhancement of the evaporation
residue cross section. Thus, an investigation of the experimental
feasibility of obtaining neutron-rich superheavy elements is an
extremely urgent subject.

\begin{figure}
\centerline{
\includegraphics[height=0.40\textheight]{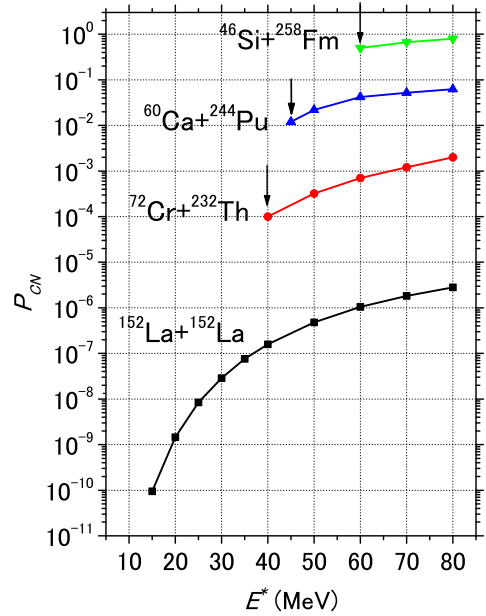}}
  \caption{Fusion probabilities for each ideal system leading the
compound nucleus $^{304}114$, which are calculated by the
three-dimensional Langevin equation. For the reaction
$^{152}$La+$^{152}$La, it is calculated by the one-dimensional
Smolcouski equation. The arrows denote the corresponding Bass
potential barriers \cite{bass74}. }
\end{figure}

\begin{figure}
\centerline{
\includegraphics[height=0.40\textheight]{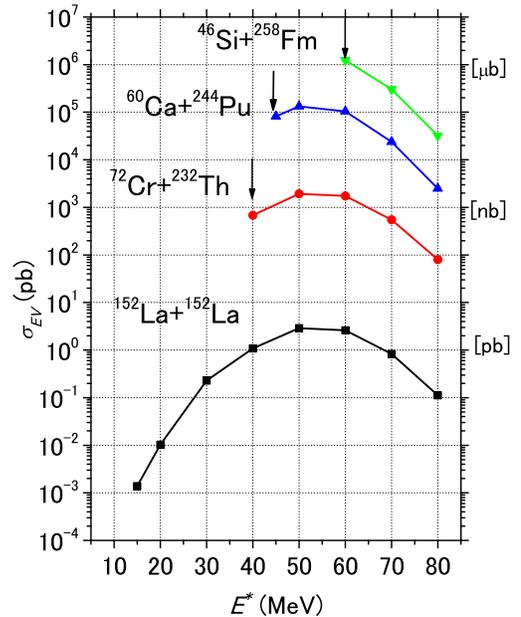}}
  \caption{Excitation function of the evaporation
residue cross section for each reaction forming the nuclei with
$Z=114$. }
\end{figure}



In summary, using the three-dimensional Langevin equation for the
fusion process and the one-dimensional Smoluchowski equation for the
survival process on the basis of our previous works
\cite{arit97,arit99,ari04}, we investigated the possibility of
synthesizing the doubly magic superheavy nucleus $^{298}114_{184}$.
Because of the neutron emissions, we must generate more neutron-rich
compound nuclei. The compound nucleus $^{304}114$ has two advantages
to achieving a high survival probability. First, because of small
neutron separation energy and rapid cooling, the shell correction
energy recovers quickly. Secondly, owing to neutron emissions, the
neutron number of the nucleus approaches that of the double closed
shell. Because of these two effects, the excitation function of the
survival probability of $^{304}114$ has a flat region around
$E^{*}=20 \sim 50$ MeV. These properties lead to a rather high
evaporation reside cross section. As a more realistic model, we plan
to take into account the emission of the charged particles from the
compound nucleus.

Although the combinations of stable nuclei cannot yield such
neutron-rich nuclei as $Z=114$ and $N>184$, we hope to make use of
secondary beams in the future. We believe, the mechanism that we
discussed here can inspire new experimental studies on the synthesis
of superheavy elements. Also, such a mechanism is very interesting
and can be applied to any system that has the same properties, small
neutron separation energy and slightly larger neutron number than
the closed shell.

The author is grateful to Professor M.~Ohta, Professor T.~Wada,
Professor Yu.~Ts.~Oganessian, Professor M.G.~Itkis, Professor
V.I.~Zagrebaev and Professor F.~Hanappe for their helpful
suggestions and valuable discussion throughout the present work. The
authors thank Dr. S.~Yamaji and his collaborators, who developed the
calculation code for potential energy with two-center
parameterization. This work has been in part supported by INTAS
projects 03-01-6417.


\bibliographystyle{aipproc}   

\begin{thebibliography}{9}

\bibitem{ogan85}Yu.Ts.~Oganessian and Y.A.~Lazarev, {\em Treatise on Heavy-Ion
Science} ed. by D.A.~Bromley (Plenum, 1985), p.3.
\bibitem{munz88}G.~M\"unzenberg, Rep. Prog. Phys. {\bf 51}, 57 (1988).
\bibitem{myer66}W.D.~Myers and W.J.~Swiatecki, Nucl. Phys. {\bf 81}, 1 (1966).
\bibitem{armb85}P.~Armbruster, Ann. Rev. Nucl. Sci. {\bf 35}, 135 (1985).
\bibitem{ogan99}Yu.Ts.~Oganessian {\em et al.}, Phys. Rev. Lett.
{\bf 83}, 3154 (1999).
\bibitem{swia81}W.J.~Swiatecki, Phys. Scripta {\bf 24}, 113 (1981);
W.J.~Swiatecki, Nucl. Phys. A {\bf376}, 275 (1982).
\bibitem{bloc86}J.P.~Blocki, H.~Feldmeier and W.J.~Swiatecki, Nucl.
Phys. A {\bf 459}, 145 (1986).
%
\bibitem{bjor82}S.~Bj\o rnholm and W. J.~Swiatecki, Nucl. Phys. A {\bf 391}, 471
(1982).
\bibitem{agui89}C.E.~Aguiar, V.C.~Barbosa and R.~Donangelo, Nucl. Phys. A {\bf 491}, 301 (1989).
\bibitem{wada93}T.~Wada, Y.~Abe and N.~Carjan, Phys. Rev. Lett. {\bf 70}, 3538 (1993).
\bibitem{toku99}T.~Tokuda, T.~Wada and M.~Ohta, Prog. Theor. Phys. {\bf
101}, 607 (1999).
\bibitem{ari04}Y.~Aritomo and M.~Ohta, Nucl. Phys. A {\bf 744}, 3 (2004).
\bibitem{arit97}Y.~Aritomo, T.~Wada, M.~Ohta and Y.~Abe, Phys. Rev. C {\bf 55}, R1011 (1997).
\bibitem{arit99}Y.~Aritomo, T.~Wada, M.~Ohta and Y.~Abe, Phys. Rev. C {\bf 59}, 796 (1999).
\bibitem{zagr01}V.I.~Zagrebaev, Phys. Rev. C {\bf 64}, 034606 (2001).
\bibitem{maru72}J.~Maruhn and W.~Greiner, Z. Phys. {\bf 251}, 431 (1972).
\bibitem{sato78}K.~Sato, A.~Iwamoto, K.~Harada, S.~Yamaji, and S.~Yoshida,
Z. Phys. A {\bf 288}, 383 (1978).
\bibitem{toke81}J.~T\"oke and W.J.~Swiatecki, Nucl. Phys. A {\bf 372}, 141 (1981).
\bibitem{ohta95}M.~Ohta, Y.~Aritomo, T.~Tokuda and Y.~Abe, {\em Proc. of Tours Symp.
on Nuclear Physics II }(World Scientific, Singapore, 1995) p.480.
\bibitem{igna75}A.V.~Ignatyuk, G.N.~Smirenkin and A.S.~Tishin, Sov. J. Nucl. Phys. {\bf 21}, 255 (1975).
\bibitem{moll95}P.~M\"oller, J.R.~Nix, W.D.~Myers and W.J.~Swiatecki, Atomic Data and Nuclear Data Tables {\bf 59}, 185 (1995).
\bibitem{bass74}R.~Bass, {\em Nuclear Reactions with Heavy Ions}
(Springer, 1980).


\end{thebibliography}

\end{document}